\documentclass[prl,aps,twocolumn,showpacs]{revtex4}

\usepackage{amsmath,graphicx,epsfig}
\begin{document}

\def\pd#1#2{\frac{\partial #1}{\partial #2}}

\title{\bf  Nucleation of solitary wave complexes in two-component mixture Bose-Einstein condensates}
\author {Natalia G. Berloff}
\affiliation {Department of Applied Mathematics and Theoretical
  Physics, Centre for Mathematical Sciences,
University of Cambridge,  Cambridge, CB3 0WA, United Kingdom
}
\date{23 December 2004}

\begin {abstract} Two-component mixtures of Bose-Einstein condensates
  are shown to support solitary wave complexes that move with a
  constant velocity  preserving their
  form. I obtain the families of such solitary wave complexes in
  two-dimensional two-component mixture condensates. These solutions are classified
  according to the structure of the wavefunction in each  component.
 I show that these   complexes nucleate from
  the surface of the disk  when it moves supercritically, therefore, suggesting
  a mechanism by which these waves can be obtained in condensates by a
  moving laser beam. The
  condition for such a nucleation is derived analytically. The flow
  for supercritical disk velocities is computed numerically. The process
  of a boundary layer separation with 
  emission of either vortex pairs in each
  component or a vortex pair in one component and a ``slaved wave'' in the
  other component is elucidated.
\end{abstract}
\maketitle
The systems of the coupled  nonlinear
Schr\"odinger (NLS) equations are fundamental and universal systems
that have 
been used to describe  motions in  conservative systems of weakly nonlinear
dispersive waves in continuum mechanics, plasma physics, nonlinear optics
and condensed matter. Recent
experimental  advances in multi-component
Bose-Einstein condensates (BECs) stimulated the interest in solitary
wave solutions of these equations and their production as they define the possible
excitation states that multi-component BECs can
support. Multi-component condensates have been formed by simultaneous
trapping and cooling of atoms in distinct spin or hyperfine levels
\cite{mixture1}. There is also a hope to obtain condensed mixtures of
different atomic species \cite{mixture2}. In this
respect a lot of attention has been paid to discovering the
topological solitons and defects such as  domain walls  \cite{dw} and
skyrmions (vortons) \cite{skyrmions} in the regime of phase
separation. In the regime where phases do not separate, the complete
families of three-dimensional solitary wave solutions were obtained  in
\cite{2comp}. There solutions were classified according to the
structure of the wavefunction of each component. Four basic types
were noted: (1) vortex
  rings of various radii in each component, (2) a vortex ring in
  one component coupled to a rarefaction solitary wave of the other
  component, (3) two coupled rarefaction waves, (4) either a vortex ring
  or a rarefaction pulse coupled to a localised disturbance of a very
  low momentum. The continuous families of such waves were shown in the
  momentum-energy plane for various values of the interaction
  strengths and the relative differences between the chemical potentials.

Two-dimensional solutions of the coupled NLS equations are  relevant in the view of the
experimental setting for studies of  multi-component BECs, for
instance,   
two-dimensional vortex complexes were shown to nucleate in  rotating
two-component condensates \cite{tsubota} giving rich equilibrium structures
such as triangular, square and double-core lattices and vortex
sheets. The goal of this letter is to obtain complete families of solitary wave solutions
in two-dimensional two-component mixture BECs and to find the
conditions for their creation in bulk condensates.

For two components,
described by the wave functions $\psi_1$ and $\psi_2$, with $N_1$ and
$N_2$ particles respectively,
 the system of the Gross-Pitaevskii (GP) equations (the coupled NLS equations) on the wave functions of the condensate components
become
\begin{eqnarray}
{\rm i} \hbar \frac{\partial \psi_1}{\partial t} &=&\biggl[ -\frac{\hbar^2}{2
    m_1} \nabla^2 + V_{11}|\psi_1|^2 +V_{12}|\psi_2|^2\biggr]\psi_1,\label{two1}\\
{\rm i} \hbar \frac{\partial \psi_2}{\partial t} &=&\biggl[ -\frac{\hbar^2}{2
    m_2} \nabla^2 + V_{12}|\psi_1|^2 +V_{22}|\psi_2|^2\biggr]\psi_2, 
\label{two2}
\end{eqnarray}
where $m_i$ is the mass of the atom  of the  $i$th condensate,  and the coupling constants $V_{ij}$ are proportional to scattering
lengths $a_{ij}$ via $V_{ij}=2\pi\hbar^2a_{ij}/m_{ij}$, where
$m_{ij}=m_im_j/(m_i+m_j)$ is the reduced mass. The interaction
conserves the number of atoms of the two species, so $\int
|\psi_i|^2\, dxdy = N_i.$ To study the equilibrium properties the energy
functional has to be minimized subject to constraint on conservation
of particles leading to introduction of two chemical potentials
$\mu_1=V_{11}n_{1}+ V_{12}n_{2}, \mu_2=V_{12}n_{1}+ V_{22}n_{2}$, where $n_{i}=|\psi_{i}|^2$
is the number density in equilibrium. The dispersion relation between the frequency $\omega$ and the wave
number $k$ of the linear perturbations ($\propto \exp[i{\bf k} \cdot
{\bf x} - i \omega t] $) around homogeneous states
 is obtained as
\begin{equation}
(\omega^2-\omega_1^2)(\omega^2-\omega_1^2)=\omega_{12}^4,
\label{omega}
\end{equation}
where $\omega_i^2(k)=c_i^2k^2+\hbar^2k^4/4m_i^2$ coincides with a
one-component Bogoliubov spectrum with the customary defined sound velocity
$c_i^2=n_{i}V_{ii}/m_i$  and $\omega_{12}^2 = c_{12}^2k^2$ where
$c_{12}^4=n_{1}n_{2}V_{12}^2/m_1 m_2$. The system is dynamically stable if
the spectrum 
 (\ref{omega}) is real and positive which implies that $V_{11}V_{22}
> V_{12}^2$ for stability. The acoustic branches of Eq. (\ref{omega}) are
$\omega_{\pm}\approx c_\pm
k$ with the corresponding sound velocities
$2c_\pm^2=c_1^2+c_2^2\pm \sqrt{(c_1^2-c_2^2)^2+4c_{12}^4}$.
The solitary waves I seek below are all subsonic, so their velocity
$U$ is less than $c_{-}$.

{\it Solitary waves.} I shall restrict the parameter space by letting
$m_1=m_2=m$, $V_{11}=V_{22}$, $\alpha=V_{12}/V_{11}$.
To find axisymmetric
  solitary wave solutions moving with the velocity $U$ in the positive
  $x-$direction, I  solve
\begin{eqnarray}
2{\rm i}U\frac{\partial \psi_1}{\partial x} &=& \nabla^2 \psi_1 +
(1-|\psi_1|^2-\alpha|\psi_2|^2)\psi_1\label{ugp1} \\
2{\rm i}U\frac{\partial \psi_2}{\partial x} &=& \nabla^2 \psi_2 +
(1-\alpha|\psi_1|^2-|\psi_2|^2-\Lambda^2)\psi_2,\label{ugp2}\\
&&\psi_1 \rightarrow \psi_{1\infty}, \quad \psi_2 \rightarrow \psi_{2\infty}, \quad {\rm as}
\quad |{\bf x}| \rightarrow \infty,\nonumber
\end{eqnarray}
where a dimensionless form of Eqs.~(\ref{two1}--\ref{two2}) is used, such that the
distances are measured in units of the correlation (healing) length
$\xi=\hbar/\sqrt{2m\mu_1}$,  the frequencies are measured in units $2\mu_1/\hbar$
and the absolute values of the fields $|\psi_1|^2$ and $|\psi_2|^2$ are
measured in units of particle density $\mu_1/V_{11}$. Also present in
Eqs.~(\ref{ugp1}--\ref{ugp2})  is the measure of asymmetry  between
chemical potentials $\Lambda^2=(\mu_1-\mu_2)/\mu_1$ (where we assume
that $\mu_1 > \mu_2$). The condition of dynamical stability under which
two components do not separate is $\alpha^2<1$. The values of the wave-functions of the solitary waves
at infinity in Eqs.~(\ref{ugp1}--\ref{ugp2})
are given by $\psi_{2\infty}^2=(1-\alpha-\Lambda^2)/(1-\alpha^2)$ and
$\psi_{1\infty}^2=1-\alpha \psi_{2\infty}^2$ and the critical speed of
sound is
$4c_\pm^2=\psi_{1\infty}^2+\psi_{2\infty}^2\pm\sqrt{(\psi_{1\infty}^2-\psi_{2\infty}^2)^2+4\alpha^2\psi_{1\infty}^2\psi_{2\infty}^2}$,
so that  $c_{-}>0$ implies $\Lambda^2<1-\alpha$.

\begin{figure}[t]
\caption{(colour online) The  dispersion curves of several families of the
  solitary wave solutions of Eqs.~(\ref{ugp1}--\ref{ugp2}). The solid lines show 
  three families of solutions  with $\alpha=0.5$ and
  $\Lambda^2=0.1$. The numbers next to the dots give the velocity of
  the solitary wave. The top (black) branch corresponds to
  VP-VP  complexes. The middle (green) branch
  shows $p$ vs ${\cal E}$ for VP-SW complexes and the bottom (red) branch is the
  dispersion curve of SW-VP complexes. The dashed (blue) line across the
  solid black branch shows the VP-VP complexes for $U=0.2$,
  $\Lambda^2=0.1$ as the intercoupling parameter $\alpha$ 
  increases in increments of $0.1$ from $0.1$ (top point) to $0.7$
  (bottom point). The light grey (magenta) dashed line shows the VP-VP
  complexes for $U=0.2$ and $\alpha=0.5$ with the asymmetry parameter
  $\Lambda^2$ taking values  $0.05$ (top point), $0.1, 0.2, 0.3$
  (bottom point). The insets show the plots of $z=|\psi_1(x,y)|^2$ (top)
  and $z=|\psi_2(x,y)|^2$ (bottom) for the VP-SW complex
  with $U=0.2$, $\alpha=0.5$, $\Lambda^2=0.1$.
}
\centering
\bigskip
\epsfig{figure=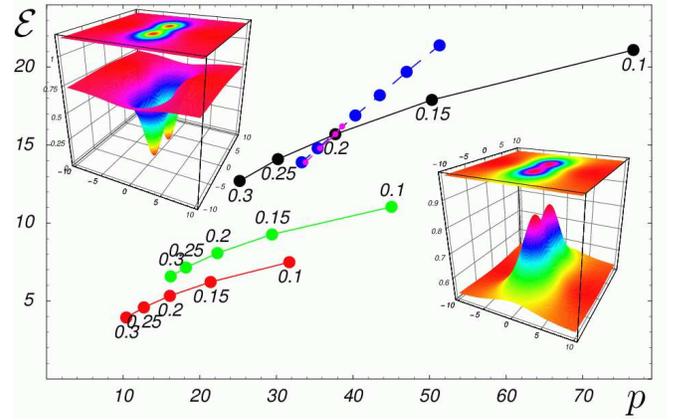, height = 2.2 in}
\label{fig1}
\end{figure}

To find each solitary wave complex,  I will fix $\alpha, \Lambda^2$, and $U$,
solve Eqs.~(\ref{ugp1}--\ref{ugp2}) by Newton-Raphson iteration procedure described
in \cite{2comp} and calculate  the momenta  ${\bf p}_i=\frac{1}{2
  {\rm i}}\int [(\psi_i^*-\psi_{i\infty})\nabla\psi_i
  -(\psi_i-\psi_{i\infty})\nabla\psi_i^*]\,dxdy$ and energy 
\begin{eqnarray}
{\cal E}&=&\frac{1}{2}\int\sum_{i=1}^2\{|\nabla \psi_i|^2
 +\frac{1}{2}(\psi_{i\infty}^2-|\psi_i|^2)^2\}
\, dxdy\nonumber \\
&&\hskip 40 pt +\frac{\alpha}{2}\int\prod_{i=1}^2(\psi_{i\infty}^2-|\psi_i|^2)\,
  dxdy\label{e1}
\end{eqnarray}
per unit length in $z$-direction.
Similarly to the three-dimensional case \cite{2comp}  we can show that 
 $U=\partial {\cal E}/\sum p_i$
  where the derivative is taken along the solitary wave
  sequence. Also, we multiply Eqs.~(\ref{ugp1}--\ref{ugp2}) by $x\partial \psi_i^*/\partial x$ and their complex conjugates
  by $x\partial \psi_i/\partial x$, integrate by parts over all
  space,  and compare the result with  (\ref{e1}) which gives ${\cal
  E}=\int\sum \left|\partial{\psi_i}/\partial {x}\right|^2\,
dx dy$. Similarly, firstly  we replace $\psi_i$ by
$\psi_i-\psi_{i\infty}$ in the first term of the
first integral of Eq.~(\ref{e1}) and integrate by parts, secondly we
 multiply Eqs.~(\ref{ugp1}--\ref{ugp2}) by $y\partial \psi_i^*/\partial y$ and
 integrate  by parts which  gives us two more integral properties that will be used
 as checks of the numerical work

\begin{eqnarray}
&& U\sum p_i=\frac{1}{2}\int \sum (\psi_{i\infty}^2-|\psi_i|^2)^2
 dxdy\nonumber \\
&&\hskip 40 pt+\alpha\int\prod (\psi_{i\infty}^2-|\psi_i|^2)\,dxdy, \label{e4}
\\ 
&&{\cal E}=\frac{1}{4}\int(1-|\psi_1|^2-\alpha|\psi_2|^2)\nonumber
\\
&&\hskip 0.3 in\times(3\psi_{1\infty}^2-\psi_{1\infty}(\psi_1+\psi_1^*)-|\psi_1|^2)\,dxdy \nonumber\\
&&\hskip 10 pt+\frac{1}{4}\int(1-\alpha|\psi_1|^2-|\psi_2|^2)\nonumber \\
&&\hskip 0.3 in\times(3\psi_{2\infty}^2-\psi_{2\infty}(\psi_2+\psi_2^*)-|\psi_2|^2)\,dxdy.
\label{p3}
\end{eqnarray}
In the limit  
$\alpha \rightarrow 0$, two components become uncoupled, in which case  the
solitary wave sequence for each component  follows the dispersion
curve of the one-component GP equation calculated in \cite{jr}. The
family of solitary waves in 2D is represented by  a pair of  point
vortices (VP) of
opposite circulation if $U_c < 0.56 c_i$. These vortices are separated by distance
$2b_i \sim U^{-1}$ for small $U$. As the velocity increases, the wave loses its
vorticity and becomes a rarefaction pulse (RP).  As $U \rightarrow c_i$
both energy and momentum per unit length approach zero and the
solutions asymptotically approach the 2D rational solution of
Kadomtsev-Petviashvili Type I equation. The sequence merges
tangentially with the phonon branch of the dispersion curve in each
of the  uncoupled components. For $\alpha \ne 0$  $c_1 \ne c_2$, so different components become RP at
different critical values of $U$ and a  variety of complexes becomes
possible. Table 1 gives  an example of various transitions from one complex
to another as the velocity $U$  increases in the system with $\alpha=0.05$ and $\Lambda^2=0.1$. 

\bigskip 

\noindent {\bf Table 1.} {\footnotesize The
velocity, $U$,
  energy, ${\cal E}$, momenta, $p_i$, and half-separations between
  centres of the 
  point vortices, $b_i$, of the
  solitary wave solutions of Eqs.~(\ref{ugp1}--\ref{ugp2}) with $\alpha=0.05$ and
  $\Lambda^2=0.1$. The sequence terminates at $U = c_{-}\approx
  0.646.$}
\medskip

\begin{tabular}{c  c c c c c c}
\hline\hline
$U$ & ${\cal E}$  & $p_1$ & $p_2$ & $b_1$ & $b_2$ & complex \\
\hline
0.40 & 14.7 & 13.8 & 12.1 & 0.915 & 0.498 & VP-VP \\
0.43 & 13.7 & 12.5 & 10.9 & 0.184 & -- & VP-RP \\
0.45 & 13.0 & 11.7& 10.2 & -- & --   & RP-RP \\
0.5 & 11.4 & 9.90 & 8.46 &  --  & --   & RP-RP \\
0.6 & 7.68 & 6.29 & 5.36 &  --  & --   & RP-RP \\
\hline \hline
\end{tabular}
\medskip

Next I will consider the cases of  more intermediate values of
intercoupling interaction strength. In addition to the  VP-VP, VP-RP, and RP-RP
complexes, there are  new classes of solitary waves that have no analog
in one-component condensates. In these complexes the disturbance of a
very low impulse in
one condensate (called ``slaved wave'' (SW)) is dragged by either VP
or RP structure of the other component. The density  of SW is maximal
where the density of either VP or RP is minimal and vise versa. For
fixed values of $\alpha$ and $\Lambda^2$ the system has three families
of solitary wave complexes: VP(RP)-VP(RP), SW-VP(RP) and VP(RP)-SW as
Fig.1 illustrates. Also,
Fig.\ref{fig1} shows the dispersion
curves  of several other
families of the solitary wave solutions in the system
 when two out of three parameters ($\alpha, \Lambda^2$ and $U$) are kept fixed. 

{\it Nucleation.} In a pioneering paper Frisch et al. \cite{frisch}
used a direct numerical simulation of 
 the one-component GP equation to show that the superflow around
a disk releases vortices from the perimeter of the disk creating a net
drag force beyond a critical velocity. The criteria for vortex
nucleation was related to the transonic transition, namely, the
vortices are created when the local speed of sound is reached
somewhere in the mainstream. The argument supporting this conclusion
is based on the observation that the hydrodynamical form of the steady
state one component GP  equation away from the disk boundary changes
its type from being elliptic to
hyperbolic if the local speed of sound is exceeded. One would expect
that somewhat similar scenario should exist for vortex nucleation in
multi-component condensates. But in two-component condensates there
are several sound speeds: $c_1, c_2, c_{12}, c_{-}$ and $c_{+}$. So
which of these values  leads to vortex nucleation? The argument based on the
criterion of the dynamical stability that the vortices should nucleate
as soon as the flow reaches the local $c_{-}$ value can be refuted by
the reference to a one-component GP equation with a nonlocal
potential that allows for the roton minimum in the dispersion curve
\cite{roton}
and therefore has two critical velocities: speed of sound $c$ and the
Landau critical velocity $v_L$, $v_L < c$. We have shown in
\cite{roton} that the nucleation of vortices is related to $c$ and not
to $v_L$. Another difficulty is that in multi-component condensates
each component has its own velocity ${\bf u}_i$. Will vortices nucleate when
the velocity of just one component reaches the criticality or is there
a more intrinsic relationship between the velocities of the components
and critical velocity? Finally, as we discovered above there exist
more than one vortex complex, so which complex nucleates at the
criticality? Next I answer these questions using a simple analytical argument and
direct numerical simulations.

By using the Madelung transformations $\psi_i=R_i \exp[{\rm i}S_i]$ in
  Eqs.~(\ref{two1}--\ref{two2}) and separating the real and imaginary parts,
  one gets the following hydrodynamical equations for the number
  density $n_i=R_i^2$ and the phase $\phi_i=\hbar S_i/m$ for the
  superflow with $\phi_i=u_{\infty}x$ as $x^2 + y^2 \rightarrow \infty$
\begin{eqnarray}
&&\pd {n_i}{t} + \nabla\cdot(n_i \nabla \phi_i)=0
  \label{hyd1}\\
&&\pd
  {\phi_i}{t}+\frac{1}{2}|\nabla\phi_i|^2-\frac{1}{2}u_{\infty}^2+\frac{V_{ii}}{m_i}(n_i-n_{i\infty})
  \nonumber \\
&&\hskip 40 pt+\frac{V_{12}}{m_i}(n_{j}-n_{j\infty})
   = \frac{\hbar^2}{2m_i^2} \frac{\nabla^2
    n_i^{1/2}}{n_i^{1/2}}, \label{hyd2}
\end{eqnarray}
where $j=2$ if $i=1$ and $j=1$ if $i=2$. We consider a stationary
flow and neglect the quantum pressure terms on the right hand side of
Eq. (\ref{hyd2}) due to our interest in mainstream flow. Since two components are
coupled through their amplitudes, the velocity vectors of two
components are parallel. We fix a point outside of the disk at which the
components move with velocities ${\bf u}_i$, introduce the local orthogonal
coordinates such that the $x-$axis  is tangent to the flow and expand
$\phi_i$ in the neighbourhood of this point as $\phi_i \approx u_i x +
\epsilon \tilde \phi_i$, where $\epsilon$ is a small parameter. To the
leading order Eqs. (\ref{hyd1}-\ref{hyd2}) become
\begin{equation}
A \tilde \phi_{ixxxx} + B \tilde \phi_{ixxyy} + C\tilde \phi_{iyyyy}=0,
\label{four}
\end{equation}
where $A=\partial (n_1 u_1)/\partial u_1 \times \partial (n_2
    u_2)/\partial u_2-u_1 u_2 \partial n_1/\partial u_2 \times
    \partial
    n_2/\partial u_1$, $B=n_2\partial (n_1 u_1)/\partial u_1+n_1\partial (n_2
    u_2)/\partial u_2$ and  $C=n_1 n_2$ where now
    $n_i=n_i(u_1,u_2)$. Note, that $B^2 \geq 4 A C$ for all $u_i$,
    therefore, Eq.~(\ref{four}) is elliptic (in the sense that it has
    no real characteristics $\partial \tilde\phi_i/\partial y + \lambda
    \partial \tilde\phi_i/\partial x=0$) if and only if $A>0$. This gives the
    criterion for the vortex nucleation: the boundary layer separation
    with nucleation of vortices takes place when 
\begin{equation}
\frac{\partial (n_1 u_1)}{\partial u_1}\frac{\partial (n_2
    u_2)}{\partial u_2}=u_1 u_2 \frac{\partial n_1}{\partial u_2} 
    \frac{\partial
    n_2}{\partial u_1},
\label{crit}
\end{equation}
somewhere in the mainstream.
In our special case of equal masses and intracomponent coupling parameters,   condition (\ref{crit}) in dimensionless units  becomes
\begin{equation}
(n_1
-2 u_1^2)(n_2 -2 u_2^2)=
4\alpha^2n_1^2n_2^2.
\label{crit2}
\end{equation}
Note that if $\alpha=0$, then the criterion (\ref{crit2}) says that
the criticality occurs when the mainstream velocity reaches the local
speed of sound $2 u_c= n_c$. If $\alpha \neq 0$ and if at criticality
$u_1 \approx u_2=u_c$, then $u_c=c_{l-}$, where $c_{l-}$ is a {\it local}
 speed of
sound defined by $4c_{l-}^2=n_1 + n_2 -\sqrt{(n_1-n_2)^2+4
  \alpha^2 n_1 n_2}$, but in general
Eq. ~(\ref{crit2}) can be satisfied when $u_1$ and $u_2$ bracket $c_{l-}$.

\begin{figure}[t!]
\caption{(colour online)  The time snapshots of $|\psi_1|^2$ (top) and
$|\psi_2|^2$ (bottom)  of the solution of  Eqs.~(\ref{ugp1}--\ref{ugp2}) with
  $-2{\rm i} \psi_{it}$ added to the left hand sides with
  $\alpha=0.5$, $\Lambda^2=0.1$
for  the flow around a disk  of radius 10 moving to the right with velocity 
$U=0.28$. The solitary wave street is seen in the wake of the
disk. The complexes were emitted in the order VP-VP, SW-VP,
VP-SW, and SW-VP complex has just got emitted from the disk boundary. For this large  supercritical velocity the
VP-VP complex is first nucleated from the surface of the disk and the
insets show this moment at an earlier time. On the main panels this
complex is in the process of splitting into the VP-SW and SW-VP complexes. 
 Only
 parts of the computational box are shown.
} 
\centering
\bigskip
\epsfig{figure=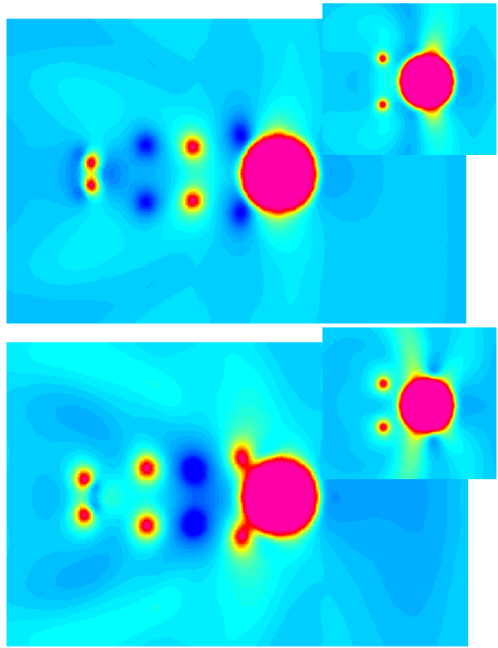, height = 3.4 in}
\label{fig2}
\end{figure}

 Direct numerical simulations of Eqs.~(\ref{two1}--\ref{two2}), with
  $-2{\rm i} \psi_{it}$ added to the left hand sides and in the frame of
  reference in which the disk is stationary  so  $U=u_\infty$,
   show  vortex
 complex  nucleation in supercritical flow around the disk. This suggests that
 these complexes  could be generated by a laser beam which moves
 supercritically in trapped condensates. 
  At subcritical velocity ($U <0.225$),
the flows of the condensates are symmetric fore and aft of the direction of 
motion,
and the disk experiences no drag. When the condition (\ref{crit2}) is
satisfied, which happens first on the disk equator where the
 velocities are maximal, the condensates evade shocks through a 
boundary layer separation.  Fig.~\ref{fig2} shows the emission of various
complexes for the disk of the radius  $10$
healing lengths that moves with supercritical velocity $U=0.28$. The disk sheds SW-VP, VP-SW and VP-VP
complexes in the order and frequency that depends on the value of the disk's velocity. These complexes
move more slowly than the disk and form a vortex wave street that trails behind 
it,
maintained by other complexes  that the disk sheds. As the velocity of the disk 
increases such a shedding becomes more and more irregular.
Each complex is born at one particular latitude within the healing layer on 
the disk. As it breaks away into the mainstream, it at first contributes
a flow that depresses the mainstream velocities on the disk below
critical. For larger values of the disk velocity ($U >0.265$), more energy is required for this depression and VP-VP complex that has
larger energy than SW-VP complexes is born first. At low supercritical
velocities SW-VP complex is born first.
As it moves further downstream however, its influence on the surface flow 
diminishes. The surface flow increases until it again reaches criticality, 
when a new  complex is nucleated and the whole sequence is
repeated. The vortex and slaved wave
street 
trailing behind the disk creates a drag on the disk that decreases as
the nearest complex moves downstream, but which is refreshed when a
new complex is born. The complexes downstream of the disk move with
 different velocities and interact among themselves. These interactions
 may lead to a transformation from one type of the complex to another;
 Fig.~\ref{fig2} shows the splitting of the VP-VP complex into SW-VP
 and VP-SW complexes. The mechanism in
 which solitary waves transfer energy from one to another was
 elucidated in \cite{pade} for a one-component condensate. 

The support from  NSF grant DMS-0104288 is 
acknowledged. 

\end{document}